\documentclass[%
 reprint,
 superscriptaddress,
 bibnotes,
 amsmath,amssymb,
 aps,
 prl,
 floatfix]{revtex4-2}
\usepackage{graphicx}
\usepackage{color}
\usepackage{dcolumn}
\usepackage{bm}
\usepackage{hyperref}
\usepackage{float}
\hypersetup{
  colorlinks   = true,
  urlcolor     = black,
  linkcolor    = black,
  citecolor   = blue
}
\usepackage{etoolbox}
\begin{document}

\title{$J=1/2$ Pseudospins and $d$-$p$ Hybridization in the Kitaev Spin Liquid Candidates Ru$X_3$ ($X$ = Cl, Br, I)}

\author{H. Gretarsson}
\affiliation{Deutsches Elektronen-Synchrotron DESY, Notkestrstra\ss e 85, D-22607 Hamburg, Germany}
\author{H. Fujihara}
\affiliation{Department of Physics, Graduate School of Science, Tohoku University, 6-3 Aramaki-Aoba, Aoba-ku, Sendai, Miyagi 980-8578, Japan}
\author{F. Sato}
\affiliation{Department of Physics, Graduate School of Science, Tohoku University, 6-3 Aramaki-Aoba, Aoba-ku, Sendai, Miyagi 980-8578, Japan}
\author{H. Gotou}
\affiliation{Institute for Solid State Physics (ISSP), University of Tokyo, Kashiwa, Chiba 277-8581, Japan}
\author{Y. Imai}
\affiliation{Department of Physics, Graduate School of Science, Tohoku University, 6-3 Aramaki-Aoba, Aoba-ku, Sendai, Miyagi 980-8578, Japan}
\author{K. Ohgushi}
\affiliation{Department of Physics, Graduate School of Science, Tohoku University, 6-3 Aramaki-Aoba, Aoba-ku, Sendai, Miyagi 980-8578, Japan}
\author{B. Keimer}
\affiliation{Max-Planck-Institut f\"{u}r Festk\"{o}rperforschung, Heisenbergstra\ss e 1, D-70569 Stuttgart, Germany}
\author{H. Suzuki}
\email[]{hakuto.suzuki@tohoku.ac.jp}
\affiliation{Frontier Research Institute for Interdisciplinary Sciences, Tohoku University, Sendai 980-8578, Japan}
\affiliation{Institute of Multidisciplinary Research for Advanced Materials (IMRAM), Tohoku University, Sendai 980-8577, Japan}
\date{\today}

\begin{abstract}
The recent synthesis of ruthenium trihalides Ru$X_3$ ($X$ = Cl, Br, I) has enlarged the set of material candidates for Kitaev spin liquid. The realization of Kitaev model necessitates the formation of $J=1/2$ pseudospins in the octahedral crystal field. We use Ru $L_3$-edge resonant inelastic x-ray scattering to investigate the evolution of multiplet structures in Ru$X_3$. We identified quasi-elastic magnetic correlations and spin-orbit transitions to the $J=3/2$ states without discernible trigonal splitting, thereby validating the $J=1/2$ description of magnetism in the Ru$X_3$ family. 
Lineshape broadening in RuI$_3$ provides evidence for its bulk metallicity in the vicinity of a bandwidth-controlled metal-insulator transition. 
Our results highlight the pivotal role of halogen $p$ orbitals in controlling the electronic properties of Ru$X_3$.
\end{abstract}

\maketitle

The exactly soluble Kitaev honeycomb model \cite{Kitaev.A_etal.Ann.-Phys.2006} embodies key concepts in modern condensed matter physics, such as quantum spin liquids, spin fractionalization, and emergent Majorana fermions. This model also holds promise as a basis for fault-tolerant quantum computation, which drives the search for its realization in physical platforms. The theoretical proposal of $t_{2g}^5$ spin-orbit Mott insulators as a solid-state platform of the Kitaev model \cite{Jackeli.G_etal.Phys.-Rev.-Lett.2009} has spurred extensive research into magnetism in 4$d$/5$d$ transition metal compounds \cite{Rau.J_etal.Annu.-Rev.-Condens.-Matter-Phys.2016,Hermanns.M_etal.Annu.-Rev.-Condens.-Matter-Phys.2018,Takagi.H_etal.Nat.-Rev.-Phys.2019}. The bond-dependent interaction in the Kitaev model can also emerge in the transition metal compounds with the high-spin $d^7$ electron configurations \cite{Liu.H_etal.Phys.-Rev.-B2018,Sano.R_etal.Phys.-Rev.-B2018,Liu.H_etal.Phys.-Rev.-Lett.2020} and in the rare-earth magnets with an odd number of 4$f$ electrons per site \cite{Li.F_etal.Phys.-Rev.-B2017,Rau.J_etal.Phys.-Rev.-B2018,Jang.S_etal.Phys.-Rev.-B2019}. These theoretical proposals have expanded the list of candidate magnets whose syntheses have been already reported \cite{Liu.H_etal.Phys.-Rev.-Lett.2020,Motome.Y_etal.J.-Phys.-Condens.-Matter2020}. The targeted synthesis effort requires experimental scrutiny to assess whether each candidate contains the essential ingredients of the Kitaev model.

Among the Kitaev spin liquid candidates, $\alpha$-RuCl$_3$ (hereafter RuCl$_3$) has been a focus of extensive research since signatures of fractionalized excitations have been identified in Raman scattering \cite{Sandilands.L_etal.Phys.-Rev.-Lett.2015}, neutron scattering \cite{Banerjee.A_etal.Nat.-Mater.2016,Banerjee.A_etal.Science2017,Do.S_etal.Nat.-Phys.2017,Banerjee.A_etal.npj-Quantum-Materials2018}, and thermal transport experiments \cite{Kasahara.Y_etal.Nature2018,Yokoi.T_etal.Science2021}. However, the zigzag antiferromagnetic order at $\sim7$ K \cite{Sears.J_etal.Phys.-Rev.-B2015} points to a certain deviation from the pure Kitaev model due to additional non-Kitaev interactions. In general, the nearest-neighbor interaction between the pseudospins on a $z$-bond, $\mathcal{H}_{ij}^{(z)}$, is expressed as \cite{Rau.J_etal.Phys.-Rev.-Lett.2014,Katukuri.V_etal.New-J.-Phys.2014}
\begin{align}
    \mathcal{H}_{ij}^{(z)}=\  &K S_i^{z}S_j^z
+J  {\bm S}_i \cdot  {\bm S}_j
+ \Gamma (S_i^xS_j^y +
S_i^yS_j^x )  \notag \\
+& \Gamma'(S_i^xS_j^z +S_i^zS_j^x+S_i^yS_j^z + S_i^zS_j^y),
\label{KH}
\end{align}
where $K$ is the bond-dependent Kitaev interaction, $J$ is Heisenberg exchange, and $\Gamma$ is the symmetric off-diagonal exchange. The additional symmetric off-diagonal exchange $\Gamma'$ becomes nonzero if the crystal field shows distortions from the cubic symmetry. For the $\gamma=x, y$ bonds, $\mathcal{H}_{ij}^{(\gamma)}$ follow from cyclic permutations of $S_i^x, S_i^y, S_i^z$. Recent resonant x-ray scattering studies \cite{Sears.J_etal.Nat.-Phys.2020,Suzuki.H_etal.Nat.-Commun.2021} have revealed the predominance of the ferromagnetic $K$ term, while the subdominant $J$ and $\Gamma$ terms are necessary to describe the zigzag order \cite{Cao.H_etal.Phys.-Rev.-B2016}. Considering the proximity of RuCl$_3$ to the Kitaev model, the fine-tuning of the magnetic Hamiltonian of RuCl$_3$ provides a viable approach to the spin-liquid state.

As the Kitaev interaction stems from the electron hopping between the nearest-neighbor $t_{2g}$ orbitals via the ligand $p$ orbitals in the edge-shared octahedra \cite{Jackeli.G_etal.Phys.-Rev.-Lett.2009}, a natural strategy is to replace Cl with heavier halogen elements, such as Br and I. Indeed, the recent synthesis of the sibling compounds RuBr$_3$ \cite{Ersan.F_etal.J.-Magn.-Magn.-Mater.2019,Imai.Y_etal.Phys.-Rev.-B2022,Prots.Y_etal.Anorg.-Allg.-Chem.2023} and RuI$_3$ \cite{Ersan.F_etal.J.-Magn.-Magn.-Mater.2019,Nawa.K_etal.J.-Phys.-Soc.-Jpn.2021,Ni.D_etal.Adv.-Mater.2022}, along with their solid solution Ru(Br$_{1-x}$I$_{x}$)$_3$ \cite{Sato.F_etal.Phys.-Rev.-B2024,Ni.D_etal.2023}, paves the way for the continuous tuning of the electronic properties. It is noteworthy that an insulator-to-metal transition occurs in Ru(Br$_{1-x}$I$_{x}$)$_3$ at the doping level of $x\sim 0.85$ \cite{Sato.F_etal.Phys.-Rev.-B2024,Ni.D_etal.2023}, placing RuI$_3$ in the metallic regime. Moreover, the introduction of polar structural asymmetry in the halogen site could stabilize an antiferromagnetic $K$ term \cite{Sugita.Y_etal.Phys.-Rev.-B2020}. Despite the large tunability of magnetism in transition metal halides, the effect of the halogen substitution on the microscopic electronic properties remains to be resolved.

\begin{figure}[ht]
  \centering
  \includegraphics[width = 8.6cm]{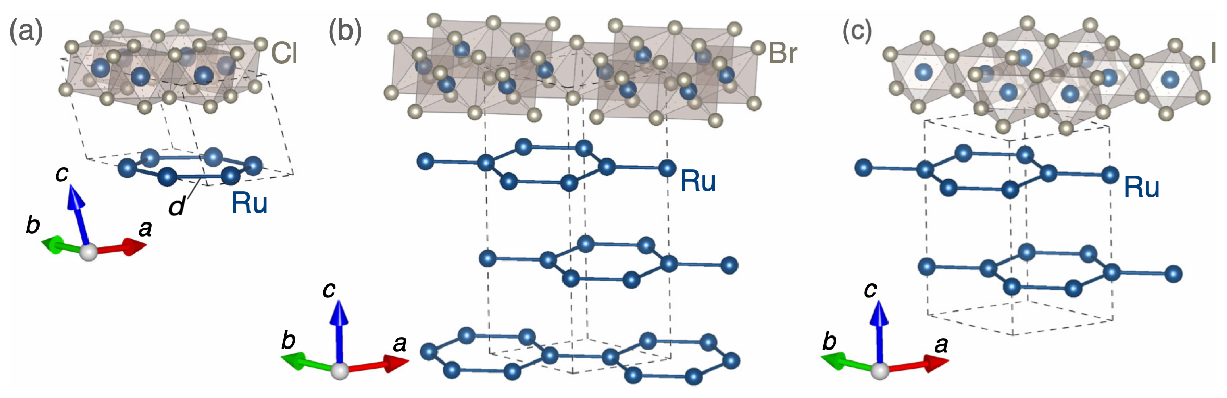}  
  \caption{\label{crys} Crystal structures of Ru$X_3$ ($X$ =  
  Cl, Br, I). (a) AlCl$_3$-type crystal structure (space group: $C2/m$) of $\alpha$-RuCl$_3$. (b) BiI$_3$-type structure $(R\overline{3})$ of RuBr$_3$. (c) Anti-Ti$_3$O-type structure $(P\overline{3}1c)$ of RuI$_3$.
  This figure is drawn using the VESTA software \cite{Momma.K_etal.J.-Appl.-Cryst.2011}.
    }
\end{figure}

In this Letter, we report a systematic Ru $L_3$-edge resonant inelastic x-ray scattering (RIXS) investigation of the multiplet structures in the ruthenium trihalides Ru$X_3$ ($X$ = Cl, Br, I). In all the Ru$X_3$ systems, we observed pseudospin $J=1/2$ quasi-elastic peaks and spin-orbit transitions to the $J=3/2$ quartet without discernible trigonal field splitting, validating the formation of the $J=1/2$ pseudospins as the basis of the magnetism of Ru$X_3$. In the metallic RuI$_3$, we identified lineshape broadening and the disappearance of the energy gap in the intersite charge continuum. This suggests that RuI$_3$ is located on the verge of a bandwidth-controlled metal-insulator transition. Multiplet analysis of the RIXS spectra reveals a systematic evolution of the crystal field parameters, due to the increase of the halogen $p$ orbital contribution to the low-energy states as X goes from Cl to I. These results establish halogen substitution as an effective tuning knob for the bulk electronic properties of Kitaev magnets.

In addition to RuCl$_3$ single crystals, we employed RuBr$_3$ and RuI$_3$ polycrystals with honeycomb lattice structures grown by high-pressure synthesis \cite{Imai.Y_etal.Phys.-Rev.-B2022,Nawa.K_etal.J.-Phys.-Soc.-Jpn.2021,Ni.D_etal.Adv.-Mater.2022}. We summarize in Fig. \ref{crys} the crystal structures of the Ru$X_3$ employed in this work.
%RuCl3
While the space group of RuCl$_3$ at low temperature is still under debate \cite{Johnson.R_etal.Phys.-Rev.-B2015,Kubota.Y_etal.Phys.-Rev.-B2015, Cao.H_etal.Phys.-Rev.-B2016,Nagai.Y_etal.Phys.-Rev.-B2020}, 
there is a consensus that RuCl$_3$ at room temperature takes an AlCl$_3$-type monoclinic crystal structure with a slightly distorted honeycomb lattice in the unit cell [$C2/m$, Fig.~\ref{crys}(a)]. The intralayer Ru-Ru bond lengths $d$ are given by $d=3.425$ \AA\ and $3.461$ \AA. 
%RuBr3
RuBr$_3$ has a BiI$_3$-type structure below room temperature without structural transition [$R\overline{3}$, Fig.~\ref{crys}(b)] \cite{Imai.Y_etal.Phys.-Rev.-B2022}, where the three regular Ru honeycomb layers with $d=3.639$ \AA\ in the unit cell form the ABCABC-type close-packed stacking sequence.
%RuI3
For RuI$_3$, different growth conditions yield two types of crystal structures \cite{Nawa.K_etal.J.-Phys.-Soc.-Jpn.2021,Ni.D_etal.Adv.-Mater.2022}.
%One is a BiI$_3$-type structure and the other 
Our RuI$_3$ crystals have an anti-Ti$_3$O-type structure [$P\overline{3}1c$, Fig.~\ref{crys}(c)] \cite{Nawa.K_etal.J.-Phys.-Soc.-Jpn.2021}, where the two regular honeycomb layers with $d=3.913$ \AA\ form the ABAB-type stacking sequence. We will discuss how the increase of $d$ affects the electronic properties below.

\begin{figure}[ht]
  \centering
  \includegraphics[width = 8.6cm]{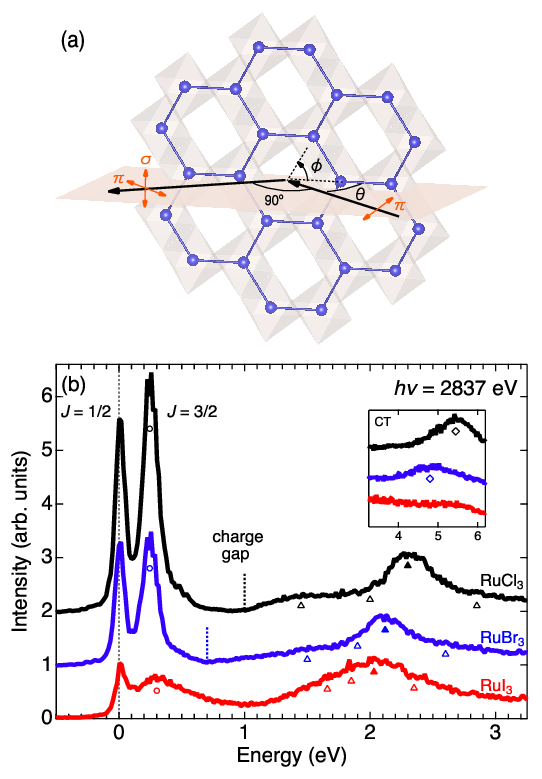}
  \caption{(a) Scattering geometry for the resonant inelastic x-ray scattering (RIXS) experiment. The incident photons were $\pi$-polarized, and the $\sigma$- and $\pi$-polarized scattered photons were collected at the scattering angle of 90$^\circ$. (b) Ru $L_3$ edge RIXS spectra of Ru$X_3$ ($X$ = Cl, Br, I) taken with the incident photons with $h\nu=$ 2837 eV. The circles indicate $J=3/2$ spin-orbit transitions. The vertical dashed lines indicate the onset energy of the intersite electron-hole continuum. Filled and open triangles indicate the main peak and shoulder structures from the crystal-field transitions to Hund's multiplets with $t_{2g}^{4}e_{g}^1$ electron configurations. The inset shows the high-energy region containing charge-transfer (CT) excitations (squares). 
  }
  \label{rixs} 
\end{figure}

Figure \ref{rixs}(a) presents the scattering geometry of the RIXS experiment,
which were conducted using the intermediate-energy RIXS (IRIXS) spectrometer at the Dynamics Beamline P01 of PETRA III, DESY \cite{Gretarsson.H_etal.J.-Synchrotron-Rad.2020}. We employed $\pi$-polarized x-ray photons with $h\nu=$ 2837 and 2839 eV, close to the Ru $L_3$ absorption edge. Scattered photons were collected at a fixed scattering angle of 90$^\circ$ using a $\mathrm{SiO_2}$ ($10\bar{2}$) diced spherical analyzer and a CCD camera. The polarizations of the outgoing photons were not analyzed. The incident x-ray beam was focused to a beam spot of 20$\times$150 $\mu$m$^{2}$ (H $\times$ V).  The exact position of zero energy loss and the energy resolution of $\Delta E \sim 80$\ meV were determined by the center and the full width at half-maximum, respectively, of the nonresonant elastic signal from silver paint deposited next to the samples. For the measurement of the RuCl$_3$ single crystal, the sample angles were set to $\theta=20^\circ$ and $\phi=0^\circ$. For the RuBr$_3$ and RuI$_3$ polycrystals, the RIXS data reflect the geometrical average over the entire range of sample angles $\theta$, $\phi$, and the tilt angle $\chi$ (not drawn). All the measurements were performed at $T=300$ K, in the paramagnetic states of Ru$X_3$.

Figure \ref{rixs}(b) shows the Ru $L_3$ RIXS spectra of Ru$X_3$ ($X$ = Cl, Br, I) taken with incident photons with $h\nu=$ 2837 eV. The spectrum for RuCl$_3$ reproduces a previous result in Ref. \cite{Suzuki.H_etal.Nat.-Commun.2021}. At low energy, the spin-orbit transitions to the $J=3/2$ states (circles) are identified in all the Ru$X_3$ compounds. The peak energies for RuCl$_3$ and RuBr$_3$ are almost identical (0.25 eV), while it shifts to a higher energy of 0.3 eV in RuI$_3$. The $J=3/2$ peaks do not exhibit discernible splitting, indicating that the trigonal distortion of the RuX$_6$ octahedra is much smaller than the energy resolution of 80 meV. The result for RuBr$_3$ agrees with Raman scattering data \cite{Choi.Y_etal.Phys.-Rev.-B2022}. The small trigonal distortion in the Ru$X_3$ family contrasts with appreciable trigonal distortion in the Ir-based Kitaev candidates \cite{Gretarsson.H_etal.Phys.-Rev.-Lett.2013}. These observations establish $J=1/2$ pseudospin physics in the Ru$X_3$ family. Moreover, for insulating RuCl$_3$ and RuBr$_3$ whose pseudospin interactions are given in Eq. (\ref{KH}), the small trigonal distortion suggests small $\Gamma^\prime$ terms. Momentum dependence of the $J=1/2$ intensity has yielded $\Gamma^\prime=0.1$ meV for RuCl$_3$ \cite{Suzuki.H_etal.Nat.-Commun.2021}. Future investigation of single crystals will allow the determination of $\Gamma^\prime$ term for RuBr$_3$.

The intermediate energy range includes broad continua originating from intersite electron-hole excitations. The onset of the continuum is located at $\sim 1$ eV for RuCl$_3$ and 0.7 eV for RuBr$_3$ (vertical dashed lines), reaffirming their Mott-insulating nature \cite{Imai.Y_etal.Phys.-Rev.-B2022}. In contrast, the continuum extends to zero energy in the metallic RuI$_3$. As RIXS is a bulk-sensitive probe, our RIXS data provide spectroscopic evidence that the metallic transport of RuI$_3$ \cite{Nawa.K_etal.J.-Phys.-Soc.-Jpn.2021,Ni.D_etal.Adv.-Mater.2022} is not caused by the grain boundary in the polycrystalline samples but reflects its intrinsic bulk metallicity. This result agrees with metallic band structures observed by angle-resolved photoemission measurements \cite{Louat.A_etal.Commun.-Phys.2024}. The charge continuum overlaps with the intra-ionic $J=3/2$ transitions, causing substantial lineshape broadening. However, the quasi-elastic $J=1/2$ excitations remain sharply peaked, indicating that the onset of the continuum is located close to zero energy. %These observations suggest that RuI$_3$ is on the verge of metal-insulator transition. 

\begin{figure*}[ht]
  \centering
  \includegraphics[width=18cm]{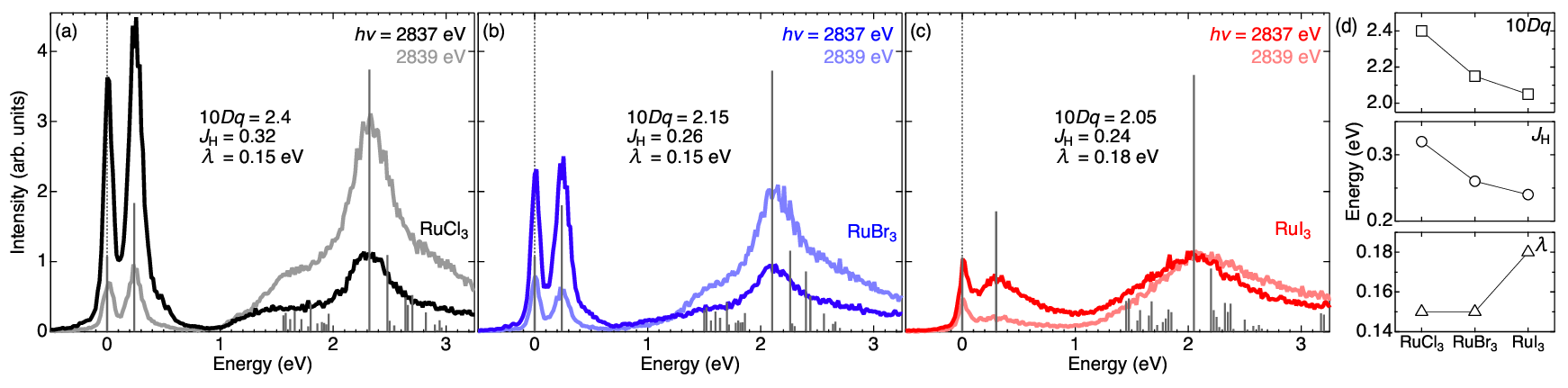}
  \caption{(a)-(c) Comparison of RIXS intensity for Ru$X_3$ taken with the $t_{2g}$ resonance ($h\nu=2837$ eV) and with the $e_g$ resonance ($h\nu=2839$ eV). The gray vertical bars indicate the theoretical RIXS intensity for the $d^5$ ionic model Hamiltonian \cite{SM}. (d) Evolution of the determined multiplet parameters 10$Dq$, $J_H$, and $\lambda$ for the Ru$X_3$ series.
  }\label{rixsfit}
\end{figure*}

On top of the continuum, one identifies intra-ionic crystal-field transitions from the $t^5_{2g}$ ground state to
the Hund's multiplets with the $t^{4}_{2g}e_{g}^1$ electron configurations (triangles). The energy of the main peaks (filled triangles) monotonically decreases as X changes from Cl to I. As this excitation energy is predominantly determined by the crystal field strength $10Dq$, the present data point to a systematic decrease of the $10Dq$ parameter. This appreciable change of the 10$Dq$ parameter [$\Delta($10$Dq)\geq 0.1$ eV, see Fig. \ref{rixsfit}(d)] is larger than that in the exfoliated nanosheets of RuCl$_3$ [$\Delta($10$Dq)\lesssim 0.05$ eV] induced by surface strain \cite{Yang.Z_etal.Phys.-Rev.-B2023}, which has been proposed as another means to control the magnetism of the Kitaev magnets \cite{Kaib.D_etal.Phys.-Rev.-B2021}. Furthermore, the energy separation of the shoulder structures (open triangles) decreases monotonically, indicating the monotonic reduction of the Hund's-rule coupling constant $J_H$ as discussed below.  

The inset of Fig. \ref{rixs}(b) shows the high-energy region of the RIXS spectra including the charge transfer (CT) excitations (squares). The peak energy in RuBr$_3$ ($\sim$ 4.8 eV) is lower than that in RuCl$_3$ ($\sim$ 5.5 eV), reflecting the lower binding energy of the Br 4$p$ orbitals in RuBr$_3$ than the Cl 3$p$ orbitals in RuCl$_3$. This leads to a stronger hybridization between the Ru 4$d$ and Br 4$p$ orbitals in RuBr$_3$, as naturally expected from the increased spatial expansion of the Br 4$p$ orbitals compared to the Cl 3$p$ orbitals. The stronger hybridization results in the suppression of CT peak intensity and the reduction of the onset of the intersite charge continuum. This trend continues in RuI$_3$. With the further enhancement of the hybridization between the Ru 4$d$ and I 5$p$ orbitals, RuI$_3$ no longer exhibits the CT peak. Instead, its high-energy spectral weight merges with the intersite charge continuum, which extends down to zero energy. The systematic enhancement of the $d$-$p$ hybridization is well captured by electronic structure calculations \cite{Nawa.K_etal.J.-Phys.-Soc.-Jpn.2021,Zhang.Y_etal.Phys.-Rev.-B2022,Kaib.D_etal.npj-Quantum-Mater.2022,Liu.L_etal.Phys.-Rev.-B2023,Samanta.S_etal.Nanomaterials2024}, which show the increase of the halogen $p$ orbital contributions in the low-energy density of states as X goes from Cl to I.

To gain more insight into the evolution of $d$-$p$ hybridization,
Figs. \ref{rixsfit}(a)-(c)  
compare the RIXS intensity for Ru$X_3$ taken with the $t_{2g}$ resonance  ($h\nu=2837$ eV) and the $e_g$ resonance ($h\nu=2839$ eV). In all the Ru$X_3$ systems, the latter condition diminishes the $J=1/2$ and $J=3/2$ transitions within the $t_{2g}^5$ sector. Concomitantly, it enhances the high-energy transitions involving the $e_g$ orbitals in the Mott-insulating RuCl$_3$ and RuBr$_3$. However, the resonance enhancement becomes less pronounced in RuBr$_3$, and the metallic RuI$_3$ yields almost comparable RIXS intensity in the high-energy region ($>1$ eV). We also observe the spectral weight shift in this region in RuI$_3$, that is, the RIXS signal is more fluorescence-like due to the itinerancy of the final states. This indicates the $e_g$ orbitals in RuI$_3$ show a large deviation from the Ru$^{3+}$ ionic character due to the strong hybridization with the I 5$p$ orbitals. Note that the ligand $p$ orbitals more sensitively affect the Ru $e_g$ orbitals than the $t_{2g}$ orbitals, as the $e_g$ orbitals point toward the halogen $p$ orbitals.

We now quantify microscopic multiplet
parameters in Ru$X_3$. For simplicity, we employ a $d^5$ ionic model Hamiltonian for the $4d$ electrons in a Ru$^{3+}$ ion. It consists of the intra-atomic Coulomb interaction terms in the Kanamori form,
$H_\mathrm{C}$ \cite{Sugano.S_etal.1970,Georges.A_etal.Annu.-Rev.-Condens.-Matter-Phys.2013}, the intra-atomic spin-orbit coupling (SOC) $H_\mathrm{SOC}$, and the cubic crystal field $H_\mathrm{cub}$ \cite{SM}. 
We employ spherical symmetry approximation of the interaction terms in $H_\mathrm{C}$, which imposes the condition $U'_{mm'} = U - 2J_{\mathrm{H},mm'}$. This approximation makes the multiplet energies from the ground state independent of $U$ in the ionic model with the fixed electron number 4$d^5$. We do not include the trigonal crystal field given the absence of discernible splitting in the $J=3/2$ transitions (Fig. \ref{rixs}). Free parameters are the crystal field strength 10$Dq$, the Hund's coupling between the $t_{2g}$ orbitals $J_H$, and the SOC $\lambda$. The RIXS transition amplitude from the ground state was calculated within the dipole approximation and fast collision approximation \cite{Ament.L_etal.Rev.-Mod.-Phys.2011,Kim.B_etal.Phys.-Rev.-B2017}, which makes the transition amplitude independent of the incident x-ray energy. Powder averaging of the RIXS transition amplitudes was performed in the calculations for the polycrystalline RuBr$_3$ and RuI$_3$ \cite{SM}.

The theoretical RIXS amplitudes for Ru$X_3$ are shown in Figs. \ref{rixsfit}(a)-(c) as gray vertical bars. Each panel also includes the optimal parameters 10$Dq$, $J_H$, and $\lambda$. Within the ionic model, the $J=3/2$ transitions are located at the energy of $\sim 3\lambda/2$, which determines the $\lambda$ parameter. For the Hund's multiplets involving the $e_g$ orbitals, the energy of the main peak fixes the 10$Dq$ parameter, and the energy separation of the shoulder structures determines $J_H$. It should be noted that the determined parameters represent ``effective'' values that incorporate the effect of the hybridization with the ligand $p$ orbitals and the interaction with the charge continuum. The parameters for RuCl$_3$ agree with those estimated in the former study \cite{Suzuki.H_etal.Nat.-Commun.2021}. For the Mott-insulating RuCl$_3$ and RuBr$_3$, the ionic model  captures the multiplet structures of the RIXS data very well. On the other hand, the agreement remains qualitative in the metallic RuI$_3$ due to the strong mixing with the charge continuum. Yet, the overall phenomenology persists, allowing the estimation of the multiplet parameters.

The evolution of these multiplet parameters in Ru$X_3$ is summarized in Fig. \ref{rixsfit}(d). First, we discuss the decreasing trend of 10$Dq$ (top panel). The larger ionic radii of the heavier halogen elements lead to the expansion of the crystal lattice and the Ru-Ru distance $d$. At the same time, the outermost halogen $p$ orbitals become more spatially extended. The decrease of the 10$Dq$ parameter reveals that the crystal lattice expansion outweighs the expansion of the $p$ orbitals, reducing the Coulomb repulsion between the Ru 4$d$ and the halogen $p$ electrons. The $J_H$ parameter also shows a decreasing trend (middle panel). This indicates that the reduction of the charge gap enhances the screening of the Coulomb interactions. In the band picture, the enhanced $d$-$p$ hybridization increases the overall bandwidth, which effectively reduces the electron correlations and induces the phase transition into the metallic state.

The well-defined $J=1/2$ magnetic correlations in the metallic RuI$_3$ point to its proximity to the metal-insulator transition. In the archetypal $J=1/2$ Mott insulator Sr$_2$IrO$_4$ \cite{Kim.J_etal.Phys.-Rev.-Lett.2012}, electron doping through La substitution at the Sr site induces a metallic state with backfolded Fermi surfaces \cite{Torre.A_etal.Phys.-Rev.-Lett.2015,Gretarsson.H_etal.Phys.-Rev.-Lett.2016_RIXS}. While the antiferromagnetic correlations become short-ranged with doping, the dispersion and intensity of the damped magnetic excitations remain analogous to the parent antiferromagnetic magnons \cite{Gretarsson.H_etal.Phys.-Rev.-Lett.2016_RIXS,Liu.X_etal.Phys.-Rev.-B2016}, allowing a ``paramagnon'' description of the magnetic dynamics akin to those in the hole-doped cuprates \cite{Le-Tacon.M_etal.Nat.-Phys.2011,Dean.M_etal.Nat.-Mater.2013,Peng.Y_etal.Phys.-Rev.-B2018}. On the other hand, the O $K$-edge RIXS spectra of the 4$d^5$ paramagnetic metal Sr$_2$RhO$_4$ lack paramagnons at low energy, although the energy scale of the $J=3/2$ excitons remains relevant \cite{Zimmermann.V_etal.npj-Quantum-Mater.2023}. The $J=1/2$ magnetic correlations in RuI$_3$ thus provide spectroscopic evidence for its proximity to a magnetically ordered phase. This is in line with the metal-insulator transition at the doping level of $x\sim 0.85$ in Ru(Br$_{1-x}$I$_{x}$)$_3$ \cite{Sato.F_etal.Phys.-Rev.-B2024,Ni.D_etal.2023}. The persisting magnetic fluctuations are likely responsible for the quasiparticle mass enhancement in the metallic region ($x>0.85$), which is identified from the increase in the Sommerfeld coefficient upon approaching the phase boundary \cite{Sato.F_etal.Phys.-Rev.-B2024}. %Furthermore, it suggests that the extended Kitaev-Heisenberg models with mobile carriers provide a good starting point for the theoretical description of the magnetism and metal-insulator transition in the Ru$X_3$ series.

The bottom panel of Fig. \ref{rixsfit}(d) shows the evolution of the $\lambda$ parameter. The insulating RuCl$_3$ and RuBr$_3$ have an identical $\lambda=0.15$ eV. On the other hand, the hardening of $J=3/2$ peak in RuI$_3$ leads to a larger $\lambda=0.18$ eV. Note here that the increased $d$-$p$ covalency reduces the orbital angular momentum of the $t_{2g}$ electrons \cite{Sugano.S_etal.1970}, which would yield the reduction of the $\lambda$ parameter. Hence, additional effects need to be considered to describe the increasing trend. For this purpose, we consider the interaction with the charge continuum and the SOC of the I 5$p$ orbitals. To clarify the former effect, we refer to the RIXS spectra of electron-doped Sr$_{2-x}$La$_x$IrO$_4$ \cite{Gretarsson.H_etal.Phys.-Rev.-Lett.2016_RIXS}. While the lineshape broadening occurs with increasing doping, the energy of the $J=3/2$ transitions remains almost identical across the insulator-metal transition. This suggests that interaction with a charge continuum alone does not induce appreciable hardening of the $J=3/2$ transitions. The increase of the $\lambda$ parameter in RuI$_3$ is thus primarily attributed to the SOC of the I 5$p$ orbitals, which is activated via the strong mixing between the Ru $t_{2g}$ and the I 5$p$ orbitals. In general, honeycomb-lattice transition metal iodides have the potential to host higher-spin Kitaev magnetism utilizing the SOC splitting of the I 5$p$ orbitals \cite{Stavropoulos.P_etal.Phys.-Rev.-Lett.2019,Stavropoulos.P_etal.Phys.-Rev.-Res.2021}. Our results, therefore, suggest the important role of the I 5$p$ orbitals in the low-energy magnetism of transition metal iodides.

In conclusion, we have presented Ru $L_3$ RIXS measurements of the multiplet structures of the Kitaev spin liquid candidates Ru$X_3$ ($X$ = Cl, Br, I). The RIXS spectra exhibit quasi-elastic magnetic correlations and spin-orbit transitions to the $J=3/2$ states, establishing the $J=1/2$ description of magnetism in the Ru$X_3$ series. We also identified the gradual reduction of the intersite charge excitation gap, evidencing the bulk metallicity of RuI$_3$. The multiplet analysis of the RIXS spectra reveals gradual suppression of electronic correlation as X goes from Cl to I, which is attributed to the increased bandwidth due to the enhanced $d$-$p$ hybridization. Our results demonstrate the crucial role of halogen $p$ orbitals in controlling the electronic properties of Ru$X_3$.

We thank R. Valent\'{i} and J. Nasu for enlightening discussions. This work was supported by Grants-in-Aid for Scientific Research from JSPS (KAKENHI) (numbers JP22K13994, JP22H00102, JP19H05823, JP19H05822, JP22H01175, JP22K18680), JST CREST (Grant No. JP19198318), and the European Research Council under Advanced Grant No. 669550 (Com4Com).
We acknowledge DESY, a member of the Helmholtz Association HGF, for the provision of experimental facilities. The RIXS experiments were carried out at the beamline P01 of PETRA III at DESY. Sample growth at ISSP was carried out under the Visiting Researcher's Program (No. 202011-MCBXG-0006 and No. 202106-MCBXG-0073). 

\bibliography{RuX3}

\end{document}